\documentclass[12pt]{elsarticle}

\usepackage[colorlinks,linkcolor=blue,citecolor=blue,urlcolor=black,bookmarks=false,hypertexnames=true]{hyperref} 

\usepackage[version=3]{mhchem}
\usepackage{amssymb}
\usepackage{color}
\usepackage{braket}
\usepackage{xspace}
\usepackage{cleveref}
\usepackage{graphicx}
\usepackage{subfigure}
\usepackage{threeparttable}
\usepackage{textcomp}
\usepackage{setspace}
\usepackage{caption}
\usepackage{comment}

\usepackage{array}
\newcolumntype{L}[1]{>{\raggedright\let\newline\\\arraybackslash\hspace{0pt}}m{#1}}
\newcolumntype{C}[1]{>{\centering\let\newline\\\arraybackslash\hspace{0pt}}m{#1}}
\newcolumntype{R}[1]{>{\raggedleft\let\newline\\\arraybackslash\hspace{0pt}}m{#1}}

\newcommand{\h}[2]{h_{{#1}}^{{#2}}}
\newcommand{\f}[2]{f_{{#1}}^{{#2}}}
\renewcommand{\d}[2]{\delta_{{#1}}^{{#2}}}
\renewcommand{\v}[2]{{v}_{{#1}}^{{#2}}}

\renewcommand{\c}[1]{a^\dagger_{#1}}
\renewcommand{\a}[1]{a_{#1}^{}}

\newcommand{\pdm}[2]{\gamma_{{#1}}^{{#2}}}

\newcommand{\om}{\omega}

\newcommand{\e}[1]{\ensuremath{\varepsilon_{#1}}}

\DeclareUnicodeCharacter{2009}{\,}

\crefname{figure}{Figure}{Figures}
\crefname{table}{Table}{Tables}
\crefname{equation}{Eq.}{Eqs.}
\crefname{section}{Section}{Sections}
\crefname{subsection}{Section}{Sections}

\usepackage{amssymb}

\allowdisplaybreaks

\journal{ }

\begin{document}

\begin{frontmatter}



\title{{\bf Multireference Perturbation Theories Based on the Dyall Hamiltonian}}
\author[label1]{Alexander~Yu.~Sokolov\corref{cor1}}
\affiliation[label1]{organization={Department of Chemistry and Biochemistry},
	addressline={The Ohio State University}, 
	city={Columbus},
	postcode={43210}, 
	state={Ohio},
	country={USA}}
\cortext[cor1]{Corresponding author e-mail address: sokolov.8@osu.edu}
	

\begin{abstract}
The concept of Dyall zeroth-order Hamiltonian [Dyall, K.\@ G.\@ {\it J.\@ Chem.\@ Phys.}, {\bf102}, 4909–4918 (1995)] has been instrumental in the development of intruder- and parameter-free multireference perturbation theories for the efficient treatment of static and dynamic correlation in molecular systems.
In this review, we discuss two theoretical approaches based on the Dyall Hamiltonian: (i) $N$-electron valence perturbation theory and (ii) multireference algebraic diagrammatic construction theory. 
We briefly cover the technical aspects behind these theories and discuss their relationship. 
We conclude with a short description of alternative approaches based on the Dyall Hamiltonian, a summary, and an outlook on future developments.
\end{abstract}




\end{frontmatter}


\section{Introduction}
\label{sec:introduction}

Multireference perturbation theories (MRPT) are attractive theoretical approaches that are capable of capturing static and dynamic electronic correlations in molecular systems with affordable computational cost.
Unfortunately, calculations using MRPT are often plagued with intruder-state problems that give rise to unphysical features in computed potential energy surfaces and compromise simulated molecular properties \cite{Evangelisti:1987p4930,Andersson:1994p391,Evangelista:2014p054109}.
The sensitivity of MRPT approaches to intruder states is closely related to the choice of zeroth-order Hamiltonian and reference many-electron wavefunctions.
For example, using the one-electron (Fock-like) zeroth-order Hamiltonians in MRPT tends to introduce severe intruder-state problems, which can be mitigated using level shift \cite{Roos:1995p215,Forsberg:1997p196} or similarity renormalization techniques \cite{Evangelista:2014p054109,Li:2015p2097} at a cost of introducing an empirical parameter. 

In 1995, Kenneth Dyall proposed the two-electron zeroth-order Hamiltonian that allows to avoid the intruder-state problems in MRPT for complete active space reference wavefunctions \cite{Dyall:1995p4909}.
Since then, the Dyall Hamiltonian has been used as the key ingredient in a variety of intruder-free MRPT methods for calculating correlation energies and molecular properties.
In this article, we review theoretical approaches based on the Dyall Hamiltonian, focusing primarily on $N$-electron valence perturbation theory (NEVPT) and multireference algebraic diagrammatic construction theory (MR-ADC), and summarize recent developments in this area.

This review is organized as follows.
We first introduce the Dyall partitioning of the Hamiltonian (\cref{sec:dyall}) and describe the technical aspects of NEVPT in its single-state and quasidegenerate formulations (\cref{sec:nevpt}).
We then give a short overview of MR-ADC and discuss the relationship of this theoretical approach with NEVPT (\cref{sec:mradc}).
Our review concludes with a short description of alternative approaches based on the Dyall Hamiltonian (\cref{sec:other}), a summary, and an outlook on future developments (\cref{sec:summary}).

%
%
%

\section{The Dyall partitioning of the electronic Hamiltonian}
\label{sec:dyall}

The success of a multireference perturbation theory relies heavily on the choice of the zeroth-order Hamiltonian $\hat{H}^{(0)}$. 
An optimal choice of $\hat{H}^{(0)}$ must ensure that the perturbation theory converges rapidly and that it produces accurate results at low perturbation orders. 
This introduces several requirements that an ideal $\hat{H}^{(0)}$ must satisfy: 
(i) it should be as close as possible to the exact Hamiltonian $\hat{H}$ so that the perturbation $\hat{V} = \hat{H} -  \hat{H}^{(0)}$ is small,
(ii) it should be diagonal or diagonally dominant when expressed in a basis of $N$-electron wavefunctions used to solve the equations of perturbation theory order by order,
and (iii) it should be chosen such that the wavefunctions from the reference multiconfigurational calculation ($\ket{\Psi^{(0)}_I}$) correspond to the eigenfunctions of this operator:
\begin{align}
	\label{eq:h_zero_eig_problem}
\hat{H}^{(0)} \ket{\Psi^{(0)}_I} = E_I^{(0)} \ket{\Psi^{(0)}_I}
\end{align}
A widely used approach to compute $\ket{\Psi^{(0)}_I}$ is complete active-space self-consistent field (CASSCF) \cite{Hinze:1973p6424,Werner:1980p5794,Roos:1980p157,Werner:1985p5053,Siegbahn:1998p2384}, which solves the full configuration problem in selected molecular orbitals (so-called active space) and variationally minimizes the energy with respect to orbital transformations.

\begin{figure*}[t!]
	\centering
	\includegraphics[width=0.4\textwidth]{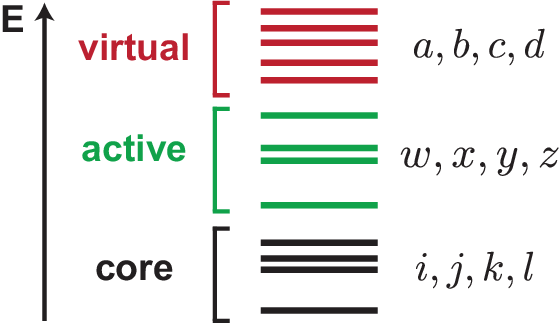}
	\captionsetup{singlelinecheck=false}
	\caption{Labeling convention for the molecular orbitals.}
	\label{fig:figure_1}
\end{figure*}

A model Hamiltonian that satisfies the above mentioned requirements for the CASSCF reference wavefunctions was proposed by Dyall in 1995 \cite{Dyall:1995p4909}. 
Expressing $\ket{\Psi^{(0)}_I}$ in the basis of core (doubly occupied), active (partially occupied), and virtual (empty) spin-orbitals $\psi_p$ with a labeling convention shown in \cref{fig:figure_1}, the Dyall Hamiltonian can be defined as:
\begin{align}
	\label{eq:h_dyall_general}
	\hat{H}^{(0)} \equiv C + \sum_{ij} \f{i}{j} \c{i}\a{j} + \sum_{ab} \f{a}{b} \c{a}\a{b} + \hat{H}_{act}
\end{align}
where $\c{p}$ is an operator that creates an electron in $\psi_p$, $\a{p}$ annihilates an electron in $\psi_p$, and $\hat{H}_{act}$ contains all active-space terms of the exact Hamiltonian $\hat{H}$
\begin{align}
	\label{eq:h_act}
	\hat{H}_{act} &= \sum_{xy}(\h{x}{y} + \sum_{i} \v{xi}{yi}) \c{x} \a{y} + \frac{1}{4} \sum_{xywz} \v{xy}{zw} \c{x} \c{y} \a{w} \a{z}
\end{align}
written using a shorthand notation for the one-electron integrals, $\h{p}{q} = \braket{p|\hat{h}|q}$, and antisymmetrized two-electron integrals, $\v{pq}{rs} = \braket{pq||rs} = \braket{pq|rs} - \braket{pq|sr}$ \cite{Helgaker:2000}.
In \cref{eq:h_dyall_general}, $\f{p}{q}$ are the matrix elements of generalized Fock operator 
\begin{align}
	\label{eq:f_gen}
	\f{p}{q} &= \h{p}{q} + \sum_{rs} \v{pr}{qs} \pdm{s}{r} 
\end{align}
that depend on the one-particle density matrix of the reference wavefunction ($\pdm{q}{p} = \braket{\Psi^{(0)}_I|\c{p}\a{q}|\Psi^{(0)}_I}$) and $C$ is a collection of constants:
\begin{align}
	\label{eq:c_term}
	C &= \sum_i \h{i}{i} + \frac{1}{2}\sum_{ij}\v{ij}{ij} - \sum_i  \f{i}{i}
\end{align}
The Dyall Hamiltonian has a simpler form when expressed in the semicanonical basis where the core and external sectors of the generalized Fock matrix are diagonal ($\f{i}{j} \rightarrow \e{i} \d{i}{j}$, $\f{a}{b} \rightarrow \e{a} \d{a}{b}$):
\begin{align}
	\label{eq:h_dyall_diagonal}
	\hat{H}^{(0)} 
	&= C + \sum_{i} \e{i} \c{i}\a{i} + \sum_{a} \e{a} \c{a} \a{a} + \hat{H}_{act}
\end{align}

Including the two-electron interaction term ($\v{xy}{zw}$) in the active-space contribution $\hat{H}_{act}$ (\cref{eq:h_act}) is the key feature that gives the Dyall Hamiltonian several attractive properties. 
First, the Dyall $\hat{H}^{(0)}$  and the exact Hamiltonian $\hat{H}$  have the same active-space contributions and the two Hamiltonians become equivalent in the full configuration interaction limit (i.e., when all orbitals are included in the active space), ensuring that the magnitude of the perturbation term $\hat{V}$ decreases as the active space grows.

Second, the Dyall form of $\hat{H}^{(0)}$ allows to satisfy \cref{eq:h_zero_eig_problem} without introducing projection operators, which need to be used to ensure this property in multireference perturbation theories with one-electron zeroth-order Hamiltonians \cite{Wolinski:1987p225,Andersson:1990p5483,Hirao:1992p374,Andersson:1992p1218,Werner:1996p645,Finley:1998p299,Ghigo:2004p142}.
To demonstrate this, we rewrite the Dyall Hamiltonian by expanding the constant term $C$ via \cref{eq:c_term} and permuting the creation and annihilation operators in the second term of \cref{eq:h_dyall_diagonal}:
\begin{align}
	\label{eq:h_dyall_diagonal_2}
	\hat{H}^{(0)} 
	&= \sum_i \h{i}{i} + \frac{1}{2}\sum_{ij}\v{ij}{ij}  - \sum_{i} \e{i} \a{i} \c{i} + \sum_{a} \e{a} \c{a} \a{a} + \hat{H}_{act}\\
	&\equiv E_{fc} - \sum_{i} \e{i} \a{i} \c{i} + \sum_{a} \e{a} \c{a} \a{a} + \hat{H}_{act}
\end{align}
Here, the first two terms of \cref{eq:h_dyall_diagonal_2} correspond to the CASSCF frozen-core energy $E_{fc}$.
Inserting this form of the Dyall Hamiltonian into \cref{eq:h_zero_eig_problem} and using the fact that $\hat{H}_{act}\ket{\Psi^{(0)}_I} = E^{(0)}_{act,I}\ket{\Psi^{(0)}_I}$, where  $E^{(0)}_{act,I}$ is the active-space contribution to the CASSCF energy $E_I^{(0)} = E_{fc} + E^{(0)}_{act,I}$, we obtain:
\begin{align}
	\label{eq:h_zero_eig_problem_2}
	\hat{H}^{(0)} \ket{\Psi^{(0)}_I} = (E_{fc} + E^{(0)}_{act,I}) \ket{\Psi^{(0)}_I}  = E_I^{(0)} \ket{\Psi^{(0)}_I}
\end{align}

Lastly, the active-space bielectronic term in \cref{eq:h_act} helps the perturbation theories based on the Dyall $\hat{H}^{(0)}$ to avoid the intruder-state problems \cite{Evangelisti:1987p4930,Andersson:1994p391,Evangelista:2014p054109,Dyall:1995p4909}, which arise due to the singular denominators in the equations. 
As discussed in Ref.\@ \citenum{Dyall:1995p4909}, the Dyall Hamiltonian may give rise to intruders only when the CASSCF electron affinity of an $N$-electron system ($E_I^{(0),N} -  E_0^{(0),N+1}$) exceeds the magnitude of its highest core orbital energy ($(E_I^{(0),N} -  E_0^{(0),N+1}) \ge \mathrm{min}(|\e{i}|)$). 
For this reason, as long as the reference CASSCF state is the ground electronic state ($\ket{\Psi^{(0)}_0}$) and the active space incorporates enough frontier orbitals such that $(E_0^{(0),N} -  E_0^{(0),N+1}) < \mathrm{min}(|\e{i}|)$, the Dyall $\hat{H}^{(0)}$ will avoid any intruder-state problems. 
Intruders may arise if the reference  wavefunction $\ket{\Psi^{(0)}_I}$ is an excited state ($I> 0$) and the electron affinity is sufficiently large to exceed $\mathrm{min}(|\e{i}|)$.
This situation is rarely observed in practice and can be avoided by incorporating the orbitals with $|\e{i}| \le (E_I^{(0),N} -  E_0^{(0),N+1})$ into the active space.

In the following, we will briefly review the theoretical approaches based on the Dyall Hamiltonian.

\section{$N$-electron valence perturbation theory}
\label{sec:nevpt}
The $N$-electron valence perturbation theory (NEVPT) proposed by Angeli, Cimrigalia, Malrieu, and co-workers \cite{Angeli:2001p10252} uses the Dyall zeroth-order Hamiltonian to incorporate dynamical correlation effects outside of the active space. 
Several variants of NEVPT methods have been developed that differ in how many states are included in the perturbation treatment (single-state vs multistate formulation), internal contraction approximations, and how the reference wavefunctions and their reduced density matrices are calculated.

\subsection{Single-state (state-specific) formulation}
\label{sec:nevpt:state_specific}

In its single-state (state-specific) form \cite{Angeli:2001p10252,Angeli:2002p9138,Angeli:2006p054108}, NEVPT is a multireference variant of Rayleigh--Schr\"odinger perturbation theory with the Dyall Hamiltonian as the zeroth-order Hamiltonian $\hat{H}^{(0)}$.
Let $\ket{\Psi_I^{(0)}}$ and $E_I^{(0)}$ be the CASSCF reference wavefunction and energy for the $I$th electronic state of interest, respectively.
The low-order corrections to the CASSCF energy have the form:
\begin{align}
	\label{eq:corr_energy_1}
	E_I^{(1)} 
	&= \braket{\Psi_I^{(0)}| \hat{{V}} |\Psi_I^{(0)}} = 0 \\
	\label{eq:corr_energy_2}
	E_I^{(2)} 
	&= \braket{\Psi_I^{(0)}|  \hat{{V}}^\dag  \frac{1}{E_I^{(0)} - \hat{{H}}^{(0)}}  \hat{{V}}  |\Psi_I^{(0)}} 
	\equiv \braket{\Psi_I^{(0)}|  \hat{{V}}  |\Psi^{(1)}_I} \\
	\label{eq:corr_energy_3}
	E_I^{(3)} 
	&= \braket{\Psi_I^{(1)}|  \hat{{V}}  |\Psi^{(1)}_I} - E_I^{(1)} \braket{\Psi_I^{(1)}|\Psi^{(1)}_I} = \braket{\Psi_I^{(1)}|  \hat{{V}}  |\Psi^{(1)}_I} 
\end{align}
The sum of $E_I^{(m)}$ up to the order $n$ define the correlation energy of {\it uncontracted} state-specific $n$th-order NEVPT (uc-NEVPT$n$).
Since the Dyall Hamiltonian satisfies \cref{eq:h_zero_eig_problem}, the first-order correction to the NEVPT$n$ energy is zero (\cref{eq:corr_energy_1}), which simplifies the expression for $E_I^{(3)}$ (\cref{eq:corr_energy_3}).

Evaluating $E_I^{(2)}$ and $E_I^{(3)}$ requires expanding the first-order wavefunction $\ket{\Psi^{(1)}_I}$ in a very large space of Slater determinants that comprise the first-order interacting space.
The size of this space grows factorially with the number of active orbitals and quadratically with the number of virtual orbitals, making the calculations of uc-NEVPT$n$ correlation energies very computationally expensive even for relatively small active spaces and modest basis sets.
Although special numerical techniques have been developed to lower the computational cost of uncontracted NEVPT calculations (see \cref{sec:nevpt:high_order_rdms} for details), $E_I^{(2)}$ and $E_I^{(3)}$ are usually evaluated by introducing {\it internal contraction approximations} that reduce the number of parameters in the first-order wavefunction $\ket{\Psi^{(1)}_I}$ and lower the computational cost. 
We will discuss the internal contraction approximations in \cref{sec:nevpt:internal_contraction}, after a short review of multistate NEVPT.

\subsection{Multistate (quasidegenerate) formulation}
\label{sec:nevpt:multistate}

While the state-specific NEVPT approach can be applied to ground or excited electronic states, it does not take into account the interaction between states when they are very close to each other in energy.
Missing this interaction in the perturbation treatment may give rise to the incorrect description of potential energy surfaces at conical intersections, avoided crossings, and in chemical systems with high density of states. 
In multistate (or quasidegenerate) formulation of NEVPT (QD-NEVPT) \cite{Angeli:2004p4043}, the coupling between nearly degenerate CASSCF states $\ket{\Psi^{(0)}_I}$ upon their perturbation is accounted for by diagonalizing the matrix of effective Hamiltonian
\begin{align}
	\label{eq:eff_H_eig_problem}
	\mathbf{{H}}_{\mathbf{eff}} \textbf{Y} = \textbf{Y}  \textbf{E}
\end{align}
following what is known as the ``diagonalize--perturb--diagonalize'' approach \cite{Zaitsevskii:1995p597, Shavitt:2008p5711}. 
In the second-order QD-NEVPT method (QD-NEVPT2) proposed by Angeli et al. \cite{Angeli:2004p4043}, the effective Hamiltonian matrix $\mathbf{{H}}_{\mathbf{eff}}$ is non-Hermitian with the following contributions at each order in perturbation theory:
\begin{align}
	\label{eq:eff_H_nosym_0}
	{H}^{(0)}_{eff,IJ} &= E^{(0)}_I \delta_{IJ} \\
	\label{eq:eff_H_nosym_1}
	{H}^{(1)}_{eff,IJ} &= \langle{\Psi^{(0)}_{I}}|\hat{{V}}|{\Psi^{(0)}_J}\rangle = 0 \\
	\label{eq:eff_H_nosym_2}
	{H}^{(2)}_{eff,IJ} &= \langle{\Psi^{(0)}_{I}}|\hat{{V}}|{\Psi^{(1)}_J}\rangle  
\end{align}
Here, the reference states $\ket{\Psi^{(0)}_I}$ are obtained from a state-averaged CASSCF calculation (SA-CASSCF) and the first-order wavefunctions $\ket{\Psi^{(1)}_I}$ are defined in \cref{eq:corr_energy_2}.

A Hermitian formulation of QD-NEVPT was developed by Sharma et al.\@ \cite{Sharma:2016p034103} using the results from canonical Van Vleck perturbation theory \cite{Kirtman:1981p798,Kirtman:2003p3890,Certain:2003p5977,Shavitt:2008p5711}.
In this approach, the effective Hamiltonian has the following form up to the third order in perturbation theory:
\begin{align}
	\label{eq:eff_H_sym_0}
	{H}^{(0)}_{eff,IJ} &= E^{(0)}_I \delta_{IJ}   \\
	\label{eq:eff_H_sym_1}
	{H}^{(1)}_{eff,IJ} &=  \langle{\Psi^{(0)}_{I}}|\hat{{V}}|{\Psi^{(0)}_J}\rangle  = 0\\
	\label{eq:eff_H_sym_2}
	{H}^{(2)}_{eff,IJ} &= \frac{1}{2} \langle{\Psi^{(0)}_{I}}|\hat{{V}}|{\Psi^{(1)}_J}\rangle +  \frac{1}{2} \langle{\Psi^{(1)}_{I}}|\hat{{V}}|{\Psi^{(0)}_J}\rangle  \\
	\label{eq:eff_H_sym_3}
	{H}^{(3)}_{eff,IJ} &=  \langle{\Psi^{(1)}_{I}}|\hat{{V}} - {H}^{(1)}_{eff,IJ}|{\Psi^{(1)}_J}\rangle  =  \langle{\Psi^{(1)}_{I}}|\hat{{V}}|{\Psi^{(1)}_J}\rangle
\end{align}
The Hermitian and non-Hermitian formulations differ in the second-order contributions to the effective Hamiltonian where \cref{eq:eff_H_sym_2} can be seen as a symmetrized version of \cref{eq:eff_H_nosym_2}.
Although the two QD-NEVPT flavors usually yield very similar results \cite{Majumder:2023p546}, the non-Hermitian variant may result in complex excitation energies while the Hermitian formulation ensures that the computed electronic energies are real-valued \cite{Lang:2020p014109}. 
The Hermiticity of QD-NEVPT effective Hamiltonian also simplifies evaluating molecular properties (such as oscillator strengths or analytic gradients, \cref{sec:nevpt:analytic_gradients_properties}), due to the symmetric form of eigenvalue problem in \cref{eq:eff_H_eig_problem}.

Diagonalizing the effective Hamiltonians defined in \cref{eq:eff_H_nosym_0,eq:eff_H_nosym_1,eq:eff_H_nosym_2} or \cref{eq:eff_H_sym_0,eq:eff_H_sym_1,eq:eff_H_sym_2,eq:eff_H_sym_3} yields the electronic energies of uncontracted QD-NEVPT$n$ approximations (uc-QD-NEVPT$n$).
Similar to state-specific NEVPT$n$, the contributions from $\ket{\Psi^{(1)}_I}$ in the QD-NEVPT$n$ effective Hamiltonian are usually evaluated by introducing internal contraction.

\subsection{Internal contraction}
\label{sec:nevpt:internal_contraction}

In the internally contracted NEVPT$n$ and QD-NEVPT$n$ methods, the first-order wavefunction for the $I$th reference state is approximated as
\begin{align}
	\label{eq:1st_order_wfn_contracted}
	\ket{\Psi^{(1)}_I} 
	\approx \hat{T}^{(1)} \ket{\Psi_I^{(0)}}
	= \sum_{\mu} t_{\mu, I}^{(1)} \hat{\tau}_\mu \ket{\Psi_I^{(0)}}
	\equiv \sum_{\mu} t_{\mu, I}^{(1)} \ket{\Phi_{\mu, I}} 
\end{align}
where $t_{\mu, I}^{(1)}$ are the first-order wavefunction parameters (amplitudes) and $\ket{\Phi_{\mu, I}}$ are the corresponding many-electron basis functions called perturbers. 
The perturber $\ket{\Phi_{\mu, I}}$  is constructed by applying the two-electron excitation operator $ \hat{\tau}_\mu$ on the reference wavefunction $\ket{\Psi_I^{(0)}}$ with {\it fixed} expansion coefficients in the basis of complete active-space determinants. 
The number of $t_{\mu, I}^{(1)}$ scales polynomially with the active space size, as opposed to the factorially scaling number of parameters in the uncontracted first-order wavefunction (\cref{eq:corr_energy_2}), making the internally contracted NEVPT$n$ and QD-NEVPT$n$ calculations more computationally efficient.

\begin{figure*}[t!]
	\includegraphics[width=1.0\textwidth]{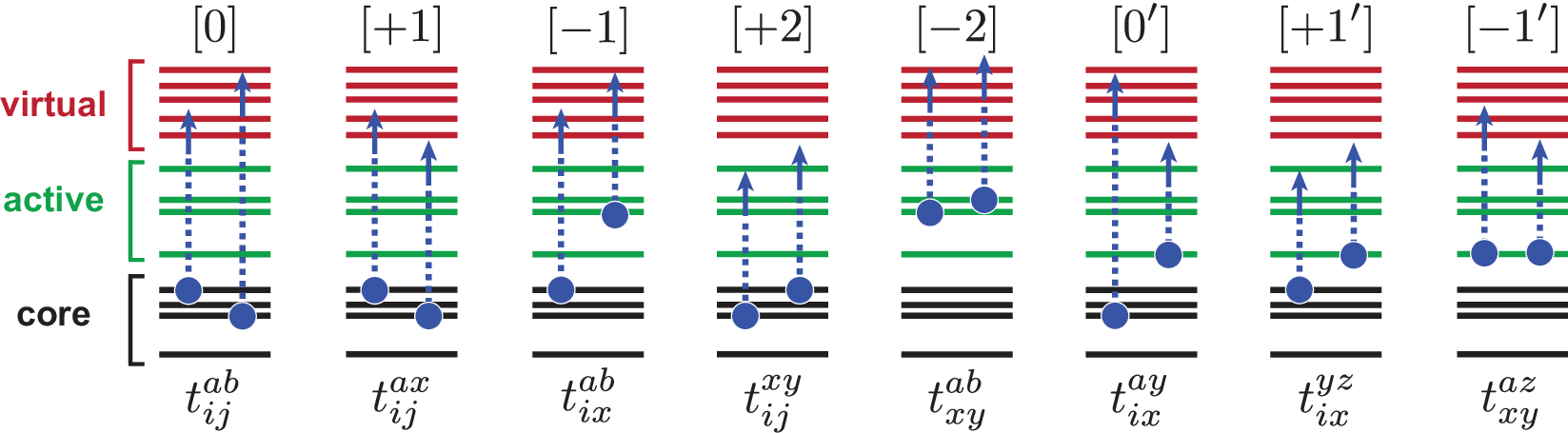}
	\captionsetup{singlelinecheck=false}
	\caption{Schematic illustration of the excitation classes described by internally contracted amplitudes $t_{\mu, I}^{(1)}$ in the first-order correlated NEVPT wavefunction $\ket{\Psi^{(1)}_I}$.}
	\label{fig:figure_2}
\end{figure*}

Each pair of $t_{\mu, I}^{(1)}$ and  $\ket{\Phi_{\mu, I}}$ belongs to one of the eight excitation classes that are labeled by the number of electrons added to or removed from the active space upon excitation: $[0]$, $[\pm 1]$, $[\pm 2]$, $[0']$, and $[\pm 1']$ (\cref{fig:figure_2}).
The $[0']$ and $[\pm 1']$ classes represent the semiinternal excitations with one of the electrons excited inside the active space.
The remaining five classes, $[0]$, $[\pm 1]$, and $[\pm 2]$, incorporate the external excitations where both electrons are transferred between different subspaces.
For each excitation class, the amplitudes $t_{\mu, I}^{(1)}$ are computed by solving a system of linear equations
\begin{align}
	\label{eq:nevpt_amplitudes}
	\sum_{\nu} K_{\mu \nu, I} t_{\nu, I}^{(1)} 
	&=-\bra{\Phi_{\mu, I}} \hat{{V}} \ket{\Psi_I^{(0)}} \\
	\label{eq:koopmans_matrix}
	K_{\mu \nu, I} &= \bra{\Phi_{\mu, I}}  (\hat{{H}}^{(0)} - E_I^{(0)})\ket{\Phi_{\nu, I}} 
\end{align}
which involves constructing and diagonalizing the matrix representation of the shifted Dyall Hamiltonian in the basis of perturbers, $K_{\mu \nu, I}$ (so-called Koopmans matrices).

Two contraction schemes have been explored within the framework of NEVPT and QD-NEVPT methods, namely: (i) {\it strong contraction} (sc) where only one perturber function is employed for each excitation class and (ii) {\it partial contraction} (pc) with multiple perturbers defined per excitation class.
The sc approach has advantage of a simpler implementation, but represents a more severe contraction approximation that introduces significant errors in the correlation energy \cite{Sokolov:2016p064102,Sivalingam:2016p054104,Sokolov:2017p244102} and orbital invariance problems in the evaluation of molecular properties \cite{Guo:2016p094111,Park:2019p5417}.
Both contraction schemes have been implemented for the NEVPT2 and QD-NEVPT2 methods \cite{Angeli:2001p10252,Angeli:2002p9138,Angeli:2004p4043}, with only strongly contracted implementation of NEVPT3 reported to date \cite{Angeli:2006p054108,Pastore:2006p445,Pastore:2006p522,Angeli:2011p184302}.
It is important to point out that the pc scheme is equivalent to full internal contraction (fic), as opposed to partial contraction in multireference configuration interaction where only a subset of excitations in the correlated wavefunction are internally contracted \cite{Siegbahn:1980p1229,Werner:1982p3144,Werner:1988p5803}.
For this reason, the pc-NEVPT and pc-QD-NEVPT methods are often abbreviated as fic-NEVPT and fic-QD-NEVPT, respectively.

\subsection{Avoiding high-order reduced density matrices}
\label{sec:nevpt:high_order_rdms}

Although internal contraction helps to reduce the computational cost of NEVPT$n$ and QD-NEVPT$n$ calculations, it can lead to problems that are not encountered in the uncontracted formulations of these methods. 
Among new challenges are the increased algebraic complexity of equations, the numerical instabilities due to nonorthogonality of perturber states, and the appearance of three- and four-particle reduced density matrices in the equations (3- and 4-RDMs). 
While the algebraic complexity can be mitigated by means of computer-aided automatic equation generation \cite{Sharma:2017p488,Chatterjee:2019p5908} and numerical instabilities can be addressed by projecting out linearly dependent perturbers, calculating many-particle reduced density matrices dominates the cost of NEVPT$n$ and QD-NEVPT$n$ calculations with 14 or more active orbitals.
In particular, the computational resources necessary to calculate the 4-RDM that appears in \cref{eq:koopmans_matrix} for the $[\pm 1']$ semiinternal excitations 
\begin{align}
	\label{eq:4rdm}
	\gamma^{wxyz}_{w'x'y'z',I} &= \braket{\Psi_I^{(0)}|\c{w}\c{x}\c{y}\c{z} \a{z'}\a{y'}\a{x'}\a{w'}|\Psi_I^{(0)}} 
\end{align}
scale as $\mathcal{O}(N_{act}^8 N_{det})$ with the number of active orbitals $N_{act}$ and the size of complete active-space Hilbert space $N_{det}$, becoming prohibitively expensive when $N_{act} \gtrsim 20$.

Several approaches have been developed to avoid calculating 4-RDM in the NEVPT$n$ and QD-NEVPT$n$ calculations that can be broadly divided into five categories:
\begin{enumerate}
	\item {\it Cumulant decomposition.} 
	The 4-RDM can be approximated in terms of lower-rank RDMs \cite{Zgid:2009p194107,Guo:2021p214111,Guo:2021p214113} by neglecting the contribution of four-particle density cumulant in its cumulant expansion \cite{Nakatsuji:1976p41,Colmenero:1993p979,Kutzelnigg:1997p432,Mazziotti:1998p4219,Kutzelnigg:1999p2800,Shamasundar:2009p174109,Kutzelnigg:2010p433}. 
	This approach suffers from the so-called false intruder problem due to the spurious eigenvalues of the approximated Koopmans matrices (\cref{eq:koopmans_matrix}) that give rise to discontinuous potential energy curves.
	The false intruders can be mitigated by introducing an imaginary level shift, which however significantly affects the accuracy of results \cite{Zgid:2009p194107}. 
	\item {\it Matrix product state compression.}
	This technique avoids internal contraction approximation and many-particle RDMs by computing the reference $\ket{\Psi_I^{(0)}}$ and uncontracted first-order $\ket{\Psi_I^{(1)}}$ wavefunctions in the matrix product state (MPS) representation \cite{Sharma:2014p111101,Sharma:2016p034103}. 
	Using density matrix renormalization group (DMRG) approach \cite{White:1992p2863,White:1999p4127,Chan:2011p465,Wouters:2014p272,Baiardi:2020p040903} allows to calculate an approximate (compressed) $\ket{\Psi_I^{(1)}}$ with a small number of MPS. 
	Increasing the number of MPS yields the NEVPT$n$ and QD-NEVPT$n$ ($n$ = 2, 3) energies that systematically converge to the fully uncontracted results \cite{Sharma:2014p111101,Sharma:2016p034103}. 
	The MPS compression can be combined with internal contraction for the energy contributions that do not depend on 4-RDM \cite{Sharma:2017p488}.
	A closely related approach utilizing the MPS compression of perturbers in sc-NEVPT2 has been developed in Ref.\@ \citenum{Roemelt:2016p204113}.
	\item {\it Imaginary time propagation.}
	Performing Laplace transformation of the energy denominator in \cref{eq:corr_energy_2} gives rise to an expression for the uncontracted NEVPT2 correlation energy as an integral in imaginary time \cite{Sokolov:2016p064102,Sokolov:2017p244102}.
	As a result, the uncontracted energies can be evaluated without computing 3- and 4-RDM, with the $\mathcal{O}(N_{act}^6 N_{det})$ cost scaling.
	The imaginary time approach has been implemented with determinant-based CASSCF and MPS-based DMRG reference wavefunctions \cite{Sokolov:2016p064102,Sokolov:2017p244102}.
	In addition, the Laplace transformation can be performed in the amplitude equations of pc-NEVPT2, enabling the calculations of fully internally contracted correlation energies without computing 4-RDM \cite{Sokolov:2018p204113}.
	\item {\it Monte Carlo techniques.}
    In this approach, stochastic sampling is employed to directly evaluate the sc-NEVPT2 correlation energy avoiding the calculation of 3- and 4-RDMs \cite{Mahajan:2019p211102,Blunt:2020p164120,Anderson:2020p054101}.
    Although the reported implementations of this technique rely on utilizing strong contraction approximation, they are embarrassingly parallel and have low computational memory requirements. 
	\item {\it Evaluating the 4-RDM terms using efficient intermediates.}
	The 4-RDM contributions can be expressed as low-order RDMs evaluated with respect to the intermediate states that are obtained by precontracting active-space two-electron integrals with the CASSCF or MPS reference wavefunctions \cite{Sokolov:2017p244102,Chatterjee:2020p6343,Kollmar:2021p234104,Majumder:2023p546}.
	Such intermediate factorization does not introduce any approximations and lowers the computational scaling to $\mathcal{O}(N_{act}^6 N_{det})$.
	Originally developed for sc-NEVPT2 \cite{Sokolov:2017p244102}, this approach has been later employed to implement pc-NEVPT2 \cite{Chatterjee:2020p6343,Kollmar:2021p234104} and pc-QD-NEVPT2 \cite{Majumder:2023p546} without 4-RDM.
\end{enumerate}
In addition to the aforementioned techniques, the cost of computing 4-RDM terms can be lowered by prescreening contributions from active-space determinants with small weights \cite{Guo:2021p214111}.
This approach, however, introduces an approximation that can affect the accuracy of results, does not completely eliminate 4-RDM, and provides modest gains in computational efficiency. 

We note that out of five categories of methods, only the intermediate factorization allows to compute the internally contracted NEVPT and QD-NEVPT correlation energies without additional approximations or user-defined parameters.
When employing the MPS compression, imaginary time propagation, or Monte Carlo techniques, the computed energies depend on the number of MPS states, the step size in time propagation, or the number of walkers in stochastic sampling.
Given the wide range of techniques available to avoid the calculation of 4-RDM, the use of cumulant approximations in internally contracted NEVPT$n$ and QD-NEVPT$n$ is not recommended.

\subsection{Increasing the active space size}
\label{sec:nevpt:large_active_spaces}

The NEVPT and QD-NEVPT calculations with conventional determinant-based CASSCF reference wavefunctions become prohibitively expensive when the size of the active space reaches $\sim$ 16 orbitals.
Calculations for larger active spaces can be performed by combining the NEVPT and QD-NEVPT methods with the DMRG or Monte Carlo techniques. 

Several DMRG implementations of uncontracted NEVPT and QD-NEVPT methods have been reported.
Sharma et al.\@ implemented NEVPT$n$ and QD-NEVPT$n$ up to $n = 3$ using the MPS compression approach \cite{Sharma:2014p111101,Sharma:2016p034103} (\cref{sec:nevpt:high_order_rdms}).
Although these methods do not require computing RDMs, the reported calculations were limited to small active spaces due to the costly MPS compression of the uncontracted first-order wavefunction.
To overcome this limitation, Ref.\@ \citenum{Sharma:2017p488} combined the MPS compression with internal contraction for the correlation energy contributions that do not depend on 4-RDM.
This implementation was used to perform computations with up to 30 active orbitals and large basis sets. 
An imaginary-time implementation of state-specific NEVPT2 with MPS has been reported in Ref.\@ \citenum{Sokolov:2017p244102} (\cref{sec:nevpt:high_order_rdms}).
This work presented uncontracted correlation energies for the chromium dimer molecule with more than 300 basis functions including 22 active orbitals and simulations of electronic states in polyene molecules with up to 24 orbitals in the active space.

The DMRG implementations of sc-NEVPT2 have been developed independently by Guo et al.\@ \cite{Guo:2016p1583} and Knecht et al.\@ \cite{Knecht:2016p244,Freitag:2017p451}.
In addition, Knecht et al.\@ reported the MPS-based implementation of pc-NEVPT2 \cite{Knecht:2016p244,Freitag:2017p451}. 
These methods required evaluating and storing 4-RDM and were used to simulate molecules with up to 22 active orbitals. 
Roemelt et al.\@ developed the implementation of sc-NEVPT2 that avoids 4-RDM by compressing perturbers in the MPS basis \cite{Roemelt:2016p204113}. 
This work carried out computations with up to 20 active orbitals.
Using the DMRG implementation of intermediate factorization (\cref{sec:nevpt:high_order_rdms}), 4-RDM-free sc-NEVPT2 calculations with up to 24 active orbitals were reported in Ref.\@ \citenum{Sokolov:2017p244102}.

The sc-NEVPT2 method has been also combined with two flavors of Monte Carlo techniques that enable large active space computations with stochastic sampling of wavefunctions and RDMs.
Anderson et al.\@ developed an implementation of sc-NEVPT2 with full configuration interaction quantum Monte Carlo and reported calculations with up to 24 active orbitals \cite{Anderson:2020p054101}.
Mahajan et al.\@ reported an efficient implementation of sc-NEVPT2 with variational Monte Carlo \cite{Mahajan:2019p211102,Blunt:2020p164120}.
This method was used to simulate the iron porphyrin and $[$Cu$_2$O$_2]^{2+}$ complexes with up to 32 orbitals in the active space.

\subsection{Extending the system size}
\label{sec:nevpt:large_basis_sets}

The NEVPT and QD-NEVPT methods can be used to perform calculations for molecules with many atoms, electrons, and orbitals.
Keeping the number of active orbitals fixed, the computational cost of NEVPT2 and QD-NEVPT2 scales as $\mathcal{O}(N^5)$ with the basis set size $N$.
For this reason, conventional implementations of these methods can be routinely applied to molecules with up to $\sim$ 1000 molecular orbitals.
Introducing the density fitting approximation \cite{Whitten:1973p4496,Dunlap:1979p3396,Vahtras:1993p514,Feyereisen:1993p359} helps to reduce the computational costs associated with handling two-electron integrals, making it possible to perform the NEVPT2 and QD-NEVPT2 calculations with $\lesssim$ 1500 basis functions using currently available computer hardware.

Several approaches have been developed to enable the NEVPT2 calculations for even larger systems.
Guo et al.\@ implemented sc- and pc-NEVPT2 with domain-based local pair natural orbitals (DLPNO) that help to reduce the computational cost by taking advantage of the local nature of electron correlation \cite{Guo:2016p094111}. 
The pc-DLPNO-NEVPT2 method was applied to molecules with more than 300 atoms and 5400 molecular orbitals recovering $\sim$ 99.9\% of pc-NEVPT2 correlation energy.
The  sc-DLPNO-NEVPT2 implementation shows large errors with respect to canonical sc-NEVPT2 due to the lack of orbital invariance.
Both pc-NEVPT2 and pc-DLPNO-NEVPT2 have been combined with the F12 approach that allows to significantly reduce the basis set incompleteness error with only a minor increase in computational cost \cite{Guo:2017p064110,Guo:2023p124120}.

The pc-DLPNO-NEVPT2 method uses nonorthogonal projected atomic orbitals as the underlying basis for constructing DLPNO.
A different approach based on non-redundant localized virtual molecular orbitals was recently developed by Uemura et al.\@ with promising numerical results \cite{Uemura:2023p154110}. 
An alternative route to developing efficient NEVPT2 implementations was proposed by Helmich--Paris and Knecht \cite{HelmichParis:2017p224101}.
This approach employs the Laplace transformation of orbital-energy denominators and formulates the NEVPT2 equations entirely in the atomic orbital basis.

Another possible strategy for large-scale NEVPT and QD-NEVPT calculations has been explored by Mitra et al.\@ who combined sc-NEVPT2 with density matrix embedding theory (sc-NEVPT2-DMET) to study the localized electronic states in extended systems \cite{Mitra:2021p11688}. 
In sc-NEVPT2-DMET, electron correlation in a small fragment of the extended system (``impurity'') is treated at the sc-NEVPT2 level of theory and is neglected elsewhere. 
As such, sc-NEVPT2-DMET is best suited for simulating the size-intensive properties of sufficiently localized electronic states, such as the excitation energies of crystalline defects. 
Applications of sc-NEVPT2-DMET to defects in diamond and magnesium oxide have been reported \cite{Haldar:2023p4273,Verma:2023p7703}.

\subsection{Analytic gradients and molecular properties}
\label{sec:nevpt:analytic_gradients_properties}

The NEVPT and QD-NEVPT methods provide access to a variety of molecular properties, which can be obtained by differentiating their energies or evaluating operator expectation values with respect to their correlated wavefunctions. 
Although the derivatives of NEVPT and QD-NEVPT energies can be evaluated using numerical differentiation techniques, such calculations are computationally expensive and can be challenging to perform. 
Park and Nishimoto implemented the analytic gradients of NEVPT2 and QD-NEVPT2 energies \cite{Nishimoto:2019p114103,Park:2019p5417,Park:2019p326,Nishimoto:2020p137219}, which allow to efficiently optimize nuclear geometries, compute adiabatic excitation energies and vibrational frequencies, and locate transition states and conical intersections on potential energy surfaces.

The NEVPT and QD-NEVPT methods have been used to compute a wide range of spin-dependent properties for open-shell molecular systems. 
These calculations require incorporating relativistic effects in the Hamiltonian and approximately solving the Dirac equation. 
Four-component relativistic sc- and pc-NEVPT2 were developed by Shiozaki et al.\@ \cite{Shiozaki:2015p4733,Reynolds:2019p1560}.
Including relativistic effects improves the accuracy of NEVPT2 for heavy element compounds and allows to calculate zero-field splitting parameters in a good agreement with experimental results. 
A less rigorous but more computationally efficient approach based on the two-component Breit--Pauli Hamiltonian and spin--orbit mean-field approximation was developed by Neese et al.\@ \cite{Neese:2005p034107,Ganyushin:2006p024103,Neese:2006p10213} and is implemented in the ORCA package for sc-NEVPT2, pc-NEVPT2, and sc-QD-NEVPT2 \cite{Neese:2022pe1606}.
This method allows to calculate the zero-field splitting energies and magnetic properties (g-tensors, electron paramagnetic resonance spectra) by perturbatively incorporating  the spin--dependent relativistic effects (spin--spin and spin--orbit coupling) \cite{Duboc:2010p10750,Maurice:2011p6229,Atanasov:2012p12324,Atanasov:2015p177,Singh:2018p4662,Lang:2019p104104,Lang:2020p014109}.
An extension of this approach to pc-QD-NEVPT2 and assessment of spin--orbit mean-field approximation was reported in Ref.\@ \citenum{Majumder:2023p546}.

\section{Multireference algebraic diagrammatic construction theory}
\label{sec:mradc}

Another theoretical approach that heavily relies on the Dyall Hamiltonian is multireference algebraic diagrammatic construction theory (MR-ADC) \cite{Sokolov:2018p204113}. 
In contrast to NEVPT and QD-NEVPT, which focus on the energies and wavefunctions of electronic states, MR-ADC uses the Dyall partitioning of the Hamiltonian to approximate the linear response function:
\begin{align}
	\label{eq:g_munu}
	G_{\mu\nu}(\omega)
	 &=  \bra{\Psi}\hat{q}_\mu(\omega - \hat{H} + E)^{-1}\hat{q}^\dag_\nu\ket{\Psi} 
	\pm \bra{\Psi}\hat{q}^\dag_\nu(\omega + \hat{H} - E)^{-1}\hat{q}_\mu\ket{\Psi} \\
	&\equiv G_{+\mu\nu}(\omega) + G_{-\mu\nu}(\omega)
\end{align}
Also known as the retarded many-body propagator, $G_{\mu\nu}(\omega)$ describes the response of a chemical system in an initial state $\ket{\Psi}$ with energy $E$ to a periodic external perturbation with frequency $\omega$ and is the central mathematical object in linear spectroscopy \cite{Fetter:1971quantum,Dickhoff:2008many,Schirmer:2018}.
The $\hat{q}^\dag_\nu$ and $\hat{q}_\mu$ called the perturbation and observable operators, respectively, determine the nature of the spectroscopic process described by $G_{\mu\nu}(\omega)$. 
For example, the charged excitations in photoelectron spectroscopy are captured by the propagator with $\hat{q}^\dag_\nu = \c{q}$ and $\hat{q}_\mu = \a{p}$ (\cref{sec:mr_adc:charged}), while the neutral excitations in UV/Vis spectroscopy are described by $\hat{q}^\dag_\nu = \c{r}\a{s} - \braket{\Psi|\c{r}\a{s}|\Psi}$ and $\hat{q}_\mu = \c{q}\a{p} - \braket{\Psi|\c{q}\a{p}|\Psi}$ (\cref{sec:mr_adc:neutral}).
The number of creation and annihilation operators in $\hat{q}^\dag_\nu$ (odd or even) determines the sign ($+$ or $-$) in \cref{eq:g_munu}.

\subsection{General formulation}
\label{sec:mr_adc:theory}

In MR-ADC, the electronic excitations and spectra are simulated by approximating $G_{\mu\nu}(\omega)$ to low order in multireference perturbation theory.
The two terms in \cref{eq:g_munu}, known as the forward ($G_{+\mu\nu}(\omega)$) and backward ($G_{-\mu\nu}(\omega)$) propagators, are expressed in the matrix form:
\begin{align}
	\label{eq:Gn_matrix}
	\mathbf{G}_{\pm}(\omega) & = \mathbf{T}_{\pm} \left(\omega \mathbf{S}_{\pm} - \mathbf{M}_{\pm}\right)^{-1} \mathbf{T}_{\pm}^{\dag}
\end{align}
where $\mathbf{M}_{\pm}$, $\mathbf{T}_{\pm}$, and $\mathbf{S}_{\pm}$ are the so-called effective Hamiltonian, effective transition moment, and overlap matrices, respectively.
Starting with the Dyall zeroth-order Hamiltonian $\hat{H}^{(0)}$ and CASSCF reference wavefunction $\ket{\Psi_I^{(0)}}$, each matrix is approximated to order $n$
\begin{align}
	\label{eq:M_pt_series}
	\mathbf{M}_{\pm} & \approx \mathbf{M}_{\pm}^{(0)} + \mathbf{M}_{\pm}^{(1)} + \ldots + \mathbf{M}_{\pm}^{(n)} \\
	\label{eq:T_pt_series}
	\mathbf{T}_{\pm} & \approx \mathbf{T}_{\pm}^{(0)} + \mathbf{T}_{\pm}^{(1)} + \ldots + \mathbf{T}_{\pm}^{(n)} \\
	\label{eq:S_pt_series}
	\mathbf{S}_{\pm} & \approx \mathbf{S}_{\pm}^{(0)} + \mathbf{S}_{\pm}^{(1)} + \ldots + \mathbf{S}_{\pm}^{(n)}
\end{align}
defining $\mathbf{G}_{\pm}(\omega)$ at the MR-ADC($n$) level of theory.
Similar to its single-reference counterpart \cite{Schirmer:1982p2395,Schirmer:1983p1237,Dreuw:2014p82,Banerjee:2023p3037}, MR-ADC ensures that $\mathbf{G}_{+}(\omega)$  and $\mathbf{G}_{-}(\omega)$ are fully decoupled from each other at any level of perturbation theory, such that their MR-ADC($n$) approximations can be constructed and computed independently.

The $\mathbf{M}_{\pm}$ and $\mathbf{T}_{\pm}$ matrices contain information about the energies and probabilities of electronic excitations, respectively, induced by the periodic perturbation.
Solving the generalized Hermitian eigenvalue problem for the effective Hamiltonian matrix $\mathbf{M}_{\pm}$
\begin{align}
	\label{eq:adc_eig_problem}
	\mathbf{M}_{\pm} \mathbf{Y}_{\pm}  = \mathbf{S}_{\pm} \mathbf{Y}_{\pm} \boldsymbol{\Omega}_{\pm}
\end{align}
yields the excitation energies $\boldsymbol{\Omega}_{\pm}$ and allows to compute the spectroscopic amplitudes
\begin{align}
	\label{eq:spec_amplitudes}
	\mathbf{X}_{\pm} = \mathbf{T}_{\pm} \mathbf{S}_{\pm}^{-1/2} \mathbf{Y}_{\pm}
\end{align}
which are related to transition intensities.
When expressed in the eigenvector basis of $\mathbf{M}_{\pm}$, the propagator takes a simple form called the spectral (or Lehmann) representation:
\begin{align}
	\label{eq:g_mr_adc}
	\mathbf{G}_{\pm}(\omega) &= \mathbf{X}_{\pm} \left(\omega - \boldsymbol{\Omega}_{\pm}\right)^{-1}  \mathbf{X}_{\pm}^\dag 
\end{align}

At each order in perturbation theory, the contributions to $\mathbf{M}_{\pm}$, $\mathbf{T}_{\pm}$, and $\mathbf{S}_{\pm}$ (\cref{eq:M_pt_series,eq:T_pt_series,eq:S_pt_series}) can be evaluated as matrix elements of the effective Hamiltonian $\tilde{H}^{(l)}$ and effective observable $\tilde{q}_{\mu}^{(k)}$ operators with respect to nonorthogonal excitations $\hat{h}_{\pm\nu}^{(l)\dagger}$
\begin{align}
	\label{eq:M+_matrix}
	M_{+\mu\nu}^{(n)} &= \sum_{klm}^{k+l+m=n} \braket{\Psi_I^{(0)}|[\hat{h}_{+\mu}^{(k)},[\tilde{H}^{(l)},\hat{h}_{+\nu}^{(m)\dagger}]]_{\pm}|\Psi_I^{(0)}} \\
	\label{eq:M-_matrix}
	M_{-\mu\nu}^{(n)} &= \sum_{klm}^{k+l+m=n} \braket{\Psi_I^{(0)}|[\hat{h}_{-\mu}^{(k)\dagger},[\tilde{H}^{(l)},\hat{h}_{-\nu}^{(m)}]]_{\pm}|\Psi_I^{(0)}}  \\
	\label{eq:T+_matrix}
	T_{+\mu\nu}^{(n)} &= \sum_{kl}^{k+l=n} \braket{\Psi_I^{(0)}|[\tilde{q}_{\mu}^{(k)},\hat{h}_{+\nu}^{(l)\dagger}]_{\pm}|\Psi_I^{(0)}} \\
	\label{eq:T-_matrix}
	T_{-\mu\nu}^{(n)} &= \sum_{kl}^{k+l=n} \braket{\Psi_I^{(0)}|[\tilde{q}_{\mu}^{(k)},\hat{h}_{-\nu}^{(l)}]_{\pm}|\Psi_I^{(0)}} \\
	\label{eq:S+_matrix}
	S_{+\mu\nu}^{(n)} &= \sum_{kl}^{k+l=n} \braket{\Psi_I^{(0)}|[\hat{h}_{+\mu}^{(k)},\hat{h}_{+\nu}^{(l)\dagger}]_{\pm}|\Psi_I^{(0)}} \\
	\label{eq:S-_matrix}
	S_{-\mu\nu}^{(n)} &= \sum_{kl}^{k+l=n} \braket{\Psi_I^{(0)}|[\hat{h}_{-\mu}^{(k)\dagger},\hat{h}_{-\nu}^{(l)}]_{\pm}|\Psi_I^{(0)}} 
\end{align}
where the sign in \cref{eq:g_munu} determines the sign in $[A,B]_{\pm} \equiv AB \pm BA$. 
The form of $\hat{h}_{\pm\nu}^{(l)\dagger}$ depends on the nature of the spectroscopic process being simulated and includes excitations inside and outside of the active space (\cref{sec:mr_adc:charged,sec:mr_adc:neutral}).

\begin{figure*}[t!]
	\includegraphics[width=1.0\textwidth]{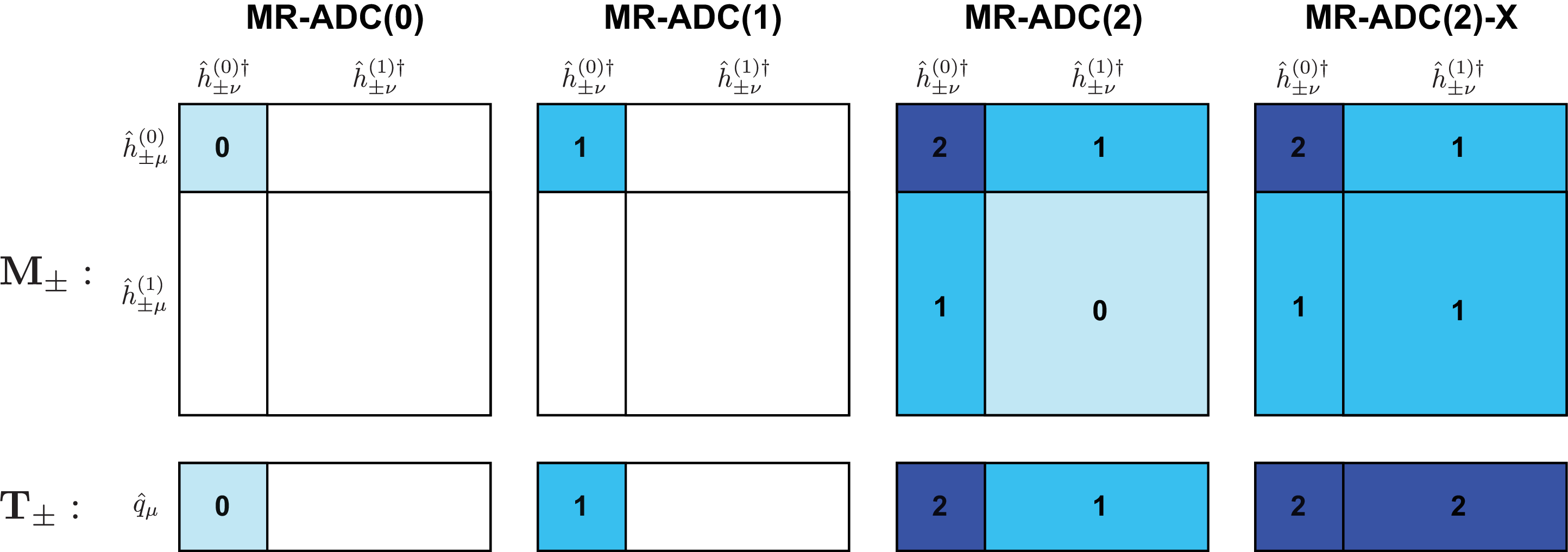}
	\captionsetup{singlelinecheck=false}
	\caption{Perturbative structures of the effective Hamiltonian ($\mathbf{M}_{\pm}$) and transition moments ($\mathbf{T}_{\pm}$) matrices in the low-order MR-ADC methods.
	Numbers denote the perturbation order to which the effective Hamiltonian and transition moments are expanded for each sector. 
	Shaded areas indicate nonzero blocks.}
	\label{fig:figure_3}
\end{figure*}

\cref{eq:M+_matrix,eq:M-_matrix,eq:T+_matrix,eq:T-_matrix} define the perturbative structure of $\mathbf{M}_{\pm}$ and $\mathbf{T}_{\pm}$ matrices where in each sector $\tilde{H}^{(l)}$ and $\tilde{q}_{\mu}^{(k)}$ are expanded to the order indicated in \cref{fig:figure_3} for the low-order MR-ADC approximations.
Up to the second order, $\tilde{H}^{(l)}$ and $\tilde{q}_{\mu}^{(k)}$ have the following contributions
\begin{align}
	\label{eq:H_bch_0}
	\tilde{H}^{(0)} &= \hat{H}^{(0)} \\
	\label{eq:H_bch_1}
	\tilde{H}^{(1)} &= \hat{V} + [\hat{H}^{(0)}, \hat{T}^{(1)} - \hat{T}^{(1)\dag}]  \\
	\label{eq:H_bch_2}
	\tilde{H}^{(2)} &= [\hat{H}^{(0)}, \hat{T}^{(2)} - \hat{T}^{(2)\dag}] + \frac{1}{2}[\hat{V} + \tilde{H}^{(1)}, \hat{T}^{(1)} - \hat{T}^{(1)\dag}] \\
	\label{eq:q_bch_0}
	\tilde{q}_\mu^{(0)} &= \hat{q}_{\mu} \\
	\label{eq:q_bch_1}
	\tilde{q}_\mu^{(1)} &= [\hat{q}_{\mu}, \hat{T}^{(1)} - \hat{T}^{(1)\dag}] \\
	\label{eq:q_bch_2}
	\tilde{q}_\mu^{(2)} &= [\hat{q}_{\mu}, \hat{T}^{(2)} - \hat{T}^{(2)\dag}] + \frac{1}{2} [[\a{p}, \hat{T}^{(1)} - \hat{T}^{(1)\dag}], \hat{T}^{(1)} - \hat{T}^{(1)\dag}]
\end{align}
where $\hat{T}^{(k)}$ is the correlation operator that defines the $k$th-order pc-NEVPT wavefunction for the reference state $\ket{\Psi_I^{(0)}}$ (e.g., \cref{eq:1st_order_wfn_contracted} for $k = 1$).

The MR-ADC(0) approximation corresponds to $\mathbf{G}_{\pm}(\omega)\approx\mathbf{G}_{\pm}^{(0)}(\omega)$ with excitation energies and transition probabilities computed using the state-specific CASSCF method with molecular orbitals and wavefunction optimized for the reference state $\ket{\Psi_I^{(0)}}$. 
Although MR-ADC(0) is equivalent to full configuration interaction when all orbitals are included in the active space, it tends to severely overestimate excitation energies for smaller active spaces since it does not account for the dynamic correlation and relaxation of non-active orbitals.
The MR-ADC(1) method incorporates some orbital relaxation effects for the excited states and is closely related to the multireference Tamm--Dancoff and random phase approximations \cite{Yeager:1979p77,Dalgaard:1980p816,Radojevic:1985p2991,Sangfelt:1998p4523}.
The accuracy of computed excitation energies and transition probabilities is significantly improved in MR-ADC(2), which adds the description of two-electron dynamic correlation and enhances the relaxation of non-active orbitals.
Additional orbital relaxation effects are included in the extended MR-ADC(2) method (MR-ADC(2)-X), which tends to be more accurate and shows the weaker dependence of results on the active space compared to MR-ADC(2).

\subsection{Charged excited states, UV and X-ray photoelectron spectra}
\label{sec:mr_adc:charged}

Choosing $\hat{q}^\dag_\nu = \c{q}$ and $\hat{q}_\mu = \a{p}$ defines the hierarchy of MR-ADC methods for electron affinities (EA-) and ionization potentials (IP-) \cite{Chatterjee:2019p5908,Chatterjee:2020p6343} as approximations for the forward and backward components of one-particle Green's function, respectively:
\begin{align}
	\label{eq:G_EA}
	G_{+pq}(\omega) & = \bra{\Psi}\a{p}(\omega - \hat{H} + E)^{-1}\c{q}\ket{\Psi} \\
	G_{-pq}(\omega) & = \bra{\Psi}\c{q}(\omega + \hat{H} - E)^{-1} \a{p}\ket{\Psi} 
\end{align}
In addition to directly calculating the energies of $(N+1)$ and $(N-1)$-electron states relative to the $N$-electron reference state $\ket{\Psi}$ (usually, the ground state), the EA/IP-MR-ADC methods can be used to compute the density of states
\begin{align}
	\label{eq:spec_function}
	A(\omega) &= -\frac{1}{\pi} \mathrm{Im} \left[ \mathrm{Tr} \, \mathbf{G}_{\pm}(\omega) \right]
\end{align}
that provides information about the intensity in photoelectron spectra.
The probability of each charged excitation ranging from zero to one is characterized by the spectroscopic factor
\begin{align}
	\label{eq:spec_factors}
	P_{\pm \alpha} = \sum_{p} |X_{\pm p\alpha}|^2
\end{align}
that can be calculated using the EA/IP-MR-ADC spectroscopic amplitudes $X_{\pm p\alpha}$ (\cref{eq:spec_amplitudes}).
Excited states with $P_{\pm \alpha} \ge 0.5$ correspond to the one-electron excitations that appear as intense features in photoelectron spectra known as the primary or quasiparticle peaks.
The weaker features with $P_{\pm \alpha} < 0.5$ represent the so-called satellite or ``shake-up'' transitions with two or more electrons being excited simultaneously.

\begin{figure*}[t!]
	\includegraphics[width=1.0\textwidth]{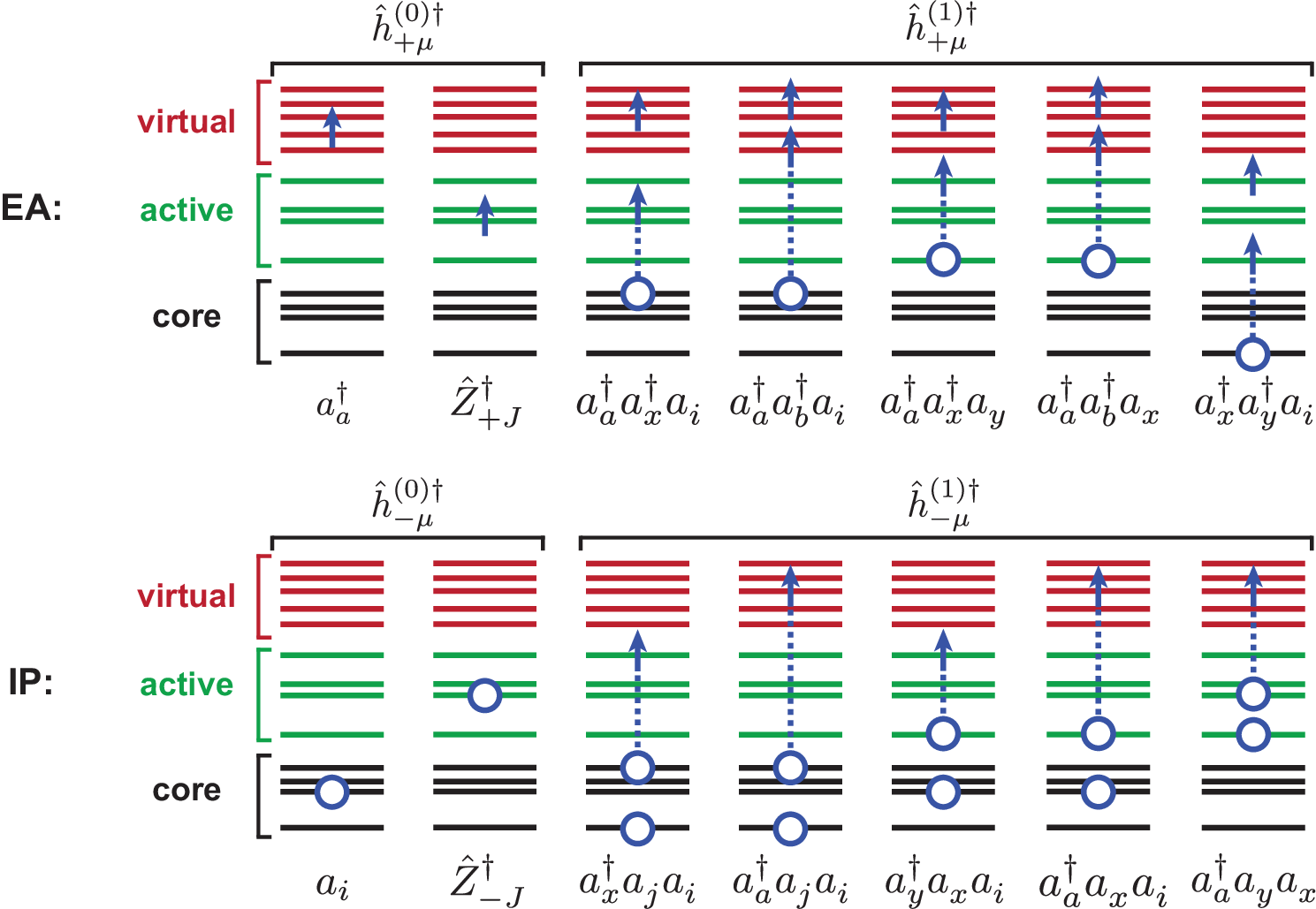}
	\captionsetup{singlelinecheck=false}
	\caption{Schematic illustration of the electron-attached and ionized states produced by acting the $\hat{h}_{\pm\mu}^{(k)\dag}$ ($k = 0, 1$) operators on the reference state $\ket{\Psi_I}$ in EA/IP-MR-ADC. 
		An arrow represents electron attachment, a circle denotes ionization, and a circle connected with an arrow denotes a single excitation. 
		The operators $\hat{Z}_{\pm J}^\dag = \ket{\Psi_J^{(0),N\pm1}}\bra{\Psi_I^{(0)}}$ incorporate all charged excitations in the active orbitals.}
	\label{fig:figure_4}
\end{figure*}

\cref{fig:figure_4} illustrates the $\hat{h}_{\pm\nu}^{(0)\dagger}$ and $\hat{h}_{\pm\nu}^{(1)\dagger}$ excitations that represent the charged excited states in the EA/IP-MR-ADC($n$) calculations ($n \le 2$).
At any level of theory, the EA/IP-MR-ADC methods incorporate all possible charged excitations in the active orbitals ($\hat{Z}_{\pm J}^\dag = \ket{\Psi_J^{(0),N\pm1}}\bra{\Psi_I^{(0)}}$) and the attachment or removal of one electron in the external ($\c{a}$) or core ($\a{i}$) orbitals, which together comprise the $\hat{h}_{\pm\nu}^{(0)\dagger}$ excitation manifold.
The double excitations in non-active orbitals ($\hat{h}_{\pm\nu}^{(1)\dagger}$) are included in EA/IP-MR-ADC($n$) with $n \ge 2$. 
In addition to capturing the satellite transitions outside of the active space, incorporating the $\hat{h}_{\pm\nu}^{(1)\dagger}$ excitations improves the treatment of orbital relaxation effects for the charged electronic states described by the $\hat{h}_{\pm\nu}^{(0)\dagger}$ operators. 

An important feature of the EA/IP-MR-ADC methods is ability to incorporate charged excitations in non-active orbitals, such as ionizing the core or inner-shell molecular orbitals with high-energy X-ray or extreme ultraviolet (XUV) light.
However, these core-level electronic transitions are deeply buried inside the spectrum of all EA/IP-MR-ADC excitations and are difficult to access using conventional iterative eigenvalue solvers, which are best suited for targeting the lowest- or highest-energy eigenstates. 
To avoid this problem, Ref.\@ \citenum{Moura:2022p4769} implemented IP-MR-ADC with core valence separation (CVS) that allows to efficiently compute the core ionization energies and X-ray photoelectron spectra (XPS)  by discarding the valence excitations from the $\hat{h}_{\pm\nu}^{(0)\dagger}$ and $\hat{h}_{\pm\nu}^{(1)\dagger}$ operator manifolds.
Although CVS neglects the interaction between the core- and valence-excited states, this coupling is expected to be small due to the large differences in their energies and spatial extent of their wavefunctions.
The CVS-IP-MR-ADC methods have been shown to correctly describe the potential energy surfaces of core-ionized states and the XPS spectra of molecules with significant multireference character \cite{Moura:2022p4769}. 

\subsection{Neutral excited states, UV/Vis and X-ray absorption spectra}
\label{sec:mr_adc:neutral}

The MR-ADC methods for neutral excitations (EE-MR-ADC) \cite{Mazin:2021p6152} approximate the linear response function with $\hat{q}^\dag_\nu = \c{r}\a{s} - \braket{\Psi|\c{r}\a{s}|\Psi}$ and $\hat{q}_\mu = \c{q}\a{p} - \braket{\Psi|\c{q}\a{p}|\Psi}$ in \cref{eq:g_munu}, known as the polarization propagator $\boldsymbol{\Pi}_{\pm}(\omega) \equiv \mathbf{G}_{\pm}(\omega)$.
In contrast to the one-particle Green's function, the forward and backward components of the polarization propagator contain the same information and are related to each other via $\mathbf{\Pi}_+^\dag(-\omega) = \mathbf{\Pi}_-(\omega)$, so it is sufficient to consider only one of them, $\mathbf{\Pi}(\omega) \equiv \mathbf{\Pi}_+(\omega)$:
\begin{align}
	\label{eq:pp}
	\Pi_{pq,rs}(\omega) = \bra{\Psi}\frac{(\c{q}\a{p} - \braket{\Psi|\c{q}\a{p}|\Psi})(\c{r}\a{s} - \braket{\Psi|\c{r}\a{s}|\Psi})}{\omega - \hat{H} + E}\ket{\Psi} 
\end{align}
Approximating $\mathbf{\Pi}(\omega)$ in the form of \cref{eq:Gn_matrix} and diagonalizing the effective Hamiltonian matrix $\mathbf{M}$ (\cref{eq:adc_eig_problem}) allows to compute the EE-MR-ADC neutral excitation energies ($\omega_k$).
The intensity of each electronic transition is described by computing the oscillator strengths
\begin{align}
	\label{eq:osc_strength}
	f_k = \frac{2}{3} \omega_k \sum_\xi \left(\sum_{pq} d_{pq\xi} X_{pqk}\right)^2
\end{align}
where $X_{pqk}$ are the EE-MR-ADC spectroscopic amplitudes and $d_{pq\xi}$ ($\xi = x, y, z$) are the matrix elements of dipole moment operator in the basis of spin-orbitals ($\vec{\mathbf{d}}$).
Expressing the EE-MR-ADC polarization propagator in its spectral form (\cref{eq:g_mr_adc}) and calculating the spectral function
\begin{equation}
	\label{eq:spectralfunction}
	T(\om) = -\frac{1}{\pi} \operatorname{Im} \left[ \operatorname{Tr} \sum_\xi\mathbf{d^{\dag}_\xi} \mathbf{\Pi}(\om) \mathbf{d}^{}_\xi \right]
\end{equation}
allows to simulate the absorption spectrum as a function of frequency $\omega$.

\begin{figure*}[t!]
	\includegraphics[width=1.0\textwidth]{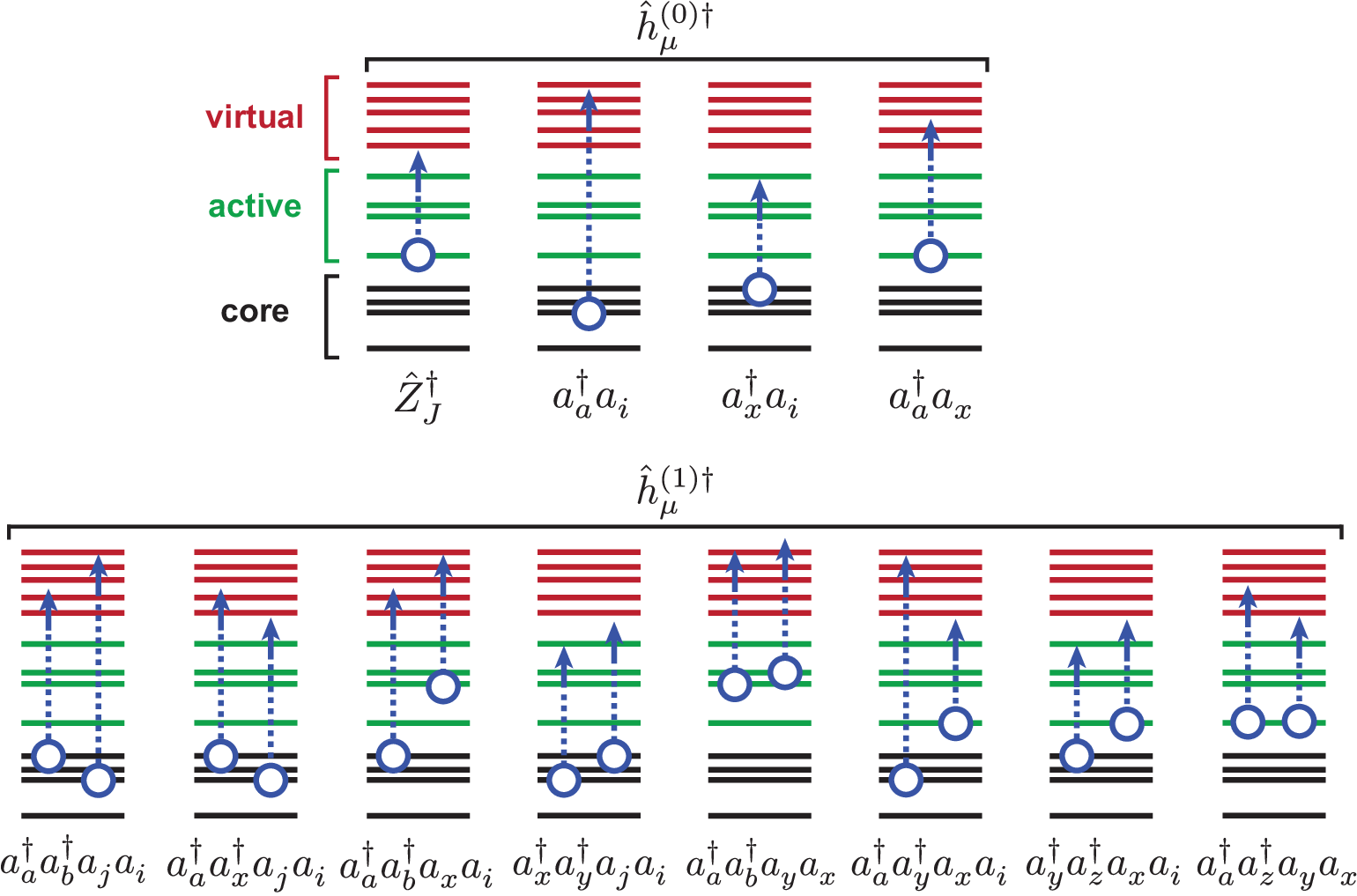}
	\captionsetup{singlelinecheck=false}
	\caption{Schematic illustration of the neutral excited states produced by acting the $\hat{h}_{\mu}^{(k)\dag}$ ($k = 0, 1$) operators on the reference state $\ket{\Psi_I}$ in EE-MR-ADC. 
		A circle connected with an arrow denotes a single excitation. 
		The operators $\hat{Z}_{J}^\dag = \ket{\Psi_J^{(0)}}\bra{\Psi_I^{(0)}}$ $(I \ne J)$ incorporate all charged excitations in the active orbitals.}
	\label{fig:figure_5}
\end{figure*}
	
The low-order EE-MR-ADC matrices $\mathbf{M}$, $\mathbf{T}$, and $\mathbf{S}$ are expressed in the basis of $\hat{h}_{\nu}^{(0)\dagger}$ and $\hat{h}_{\nu}^{(1)\dagger}$ excitations depicted in \cref{fig:figure_5}.
Starting with the reference CASSCF state $\ket{\Psi_I^{(0)}}$, the zeroth-order operators $\hat{h}_{\nu}^{(0)\dagger}$ incorporate all neutral excitations in the active space ($\hat{Z}_{J}^\dag = \ket{\Psi_J^{(0)}}\bra{\Psi_I^{(0)}}$, $J \ne I$) and a complete set of single excitations in non-active orbitals ($\c{a}\a{i}$, $\c{x}\a{i}$, and $\c{a}\a{x}$).
The double excitations outside of the active space are incorporated by the $\hat{h}_{\nu}^{(1)\dagger}$ operators.
As for EA/IP-MR-ADC, $\hat{h}_{\nu}^{(1)\dagger}$ play an important role in describing the orbital relaxation effects for the active-space singly excited electronic states captured by $\hat{h}_{\nu}^{(0)\dagger}$.

The EE-MR-ADC methods can be used to compute the energies of core-excited states and simulate the X-ray and XUV absorption spectra by introducing the CVS approximation. 
The implementation and benchmark of CVS-EE-MR-ADC(2) or CVS-EE-MR-ADC(2)-X have been reported in Ref.\@ \citenum{Mazin:2023p4991}.
While both methods produce qualitatively correct results for multireference systems, CVS-EE-MR-ADC(2)-X tends to be more accurate due to the higher-level description of orbital relaxation effects.

\subsection{Connection to NEVPT and QD-NEVPT}
\label{sec:mr_adc:connection_to_nevpt}

The MR-ADC methods have a close relationship with state-specific and quasidegenerate NEVPT. 
Here, we comment on similarities and differences between these theories.

{\it Description of excited states.} 
MR-ADC is a linear-response multistate theory that describes many excited electronic states in one calculation by starting with the state-specific reference CASSCF wavefunction for a particular electronic state and directly approximating the differences in energies and properties between the excited states and the reference state. 
This allows MR-ADC to incorporate the excitations inside and outside the active space, such as the excited states with a different number of electrons compared to that in the reference wavefunction (e.g., the ($N\pm1$)-electron states in EA/IP-MR-ADC).
In contrast, the QD-NEVPT calculations start from the state-averaged CASSCF (SA-CASSCF) calculation and approximate the energies of individual electronic states originating from the electron-number-conserving excitations within the active space. 

{\it Treatment of dynamic correlation and orbital relaxation.} 
In QD-NEVPT, the dynamic correlation effects are incorporated using a technique called multipartitioning \cite{Zaitsevskii:1995p597} where the correlated wavefunctions $\ket{\Psi_I^{(k)}}$ ($k \ge 1$) are parameterized independently for each SA-CASSCF reference state $\ket{\Psi_I^{(0)}}$. 
This approach allows to capture the differences in dynamic correlation and orbital relaxation effects between the reference states and accurately predict their relative energies.
MR-ADC takes a different strategy by performing the calculations of excited states using the orbitals and amplitudes of the correlated wavefunction determined for the reference CASSCF state $\ket{\Psi_I^{(k)}}$ ($k \ge 1$). 
The differential orbital and dynamic correlation effects are incorporated by including a large number of non-active excitations in the representation of effective Hamiltonian.
While MR-ADC($n$) tends to show lower accuracy than QD-NEVPT$n$ when the electronic structures of excited states are very different from that of the reference state, particularly at low perturbation order ($n < 2$) or when using small active spaces \cite{Chatterjee:2019p5908}, it allows to completely avoid performing the challenging SA-CASSCF calculations and evaluating the costly $\ket{\Psi_I^{(k)}}$ for more than one reference state.

{\it Treatment of quasidegeneracies and avoided crossings.} 
By employing the ``diagonalize--perturb--diagonalize'' approach, QD-NEVPT is able to accurately describe nearly degenerate electronic states when their wavefunctions have the same symmetry and strongly interact, such as in the vicinity of avoided crossing between two potential energy surfaces. 
While MR-ADC can correctly describe the strong interaction of excited-state wavefunctions, it assumes that the coupling between the reference and excited states is weak and can be treated perturbatively.
For this reason, the low-order MR-ADC methods are not expected to provide accurate results when the reference wavefunction mixes strongly with one of the excited states, unless this coupling is accurately accounted for in the reference CASSCF calculation with a large active space.

{\it Internal contraction.} 
Currently available MR-ADC methods employ full internal contraction to approximate the correlated wavefunctions $\ket{\Psi_I^{(k)}}$ ($k \ge 1$) for the reference state. 
For this reason, the reference state correlation energy computed using MR-ADC($n$) is equivalent to the state-specific pc-NEVPT$n$ correlation energy for this state.

{\it Comparison of QD-NEVPT2 and EE-MR-ADC(2) effective Hamiltonians.} 
To understand the relationship between MR-ADC and QD-NEVPT better, we compare the active-space sector of EE-MR-ADC(2) effective Hamiltonian ($M_{\mu\nu}$, $\hat{h}_{\mu}^{(0)} = \ket{\Psi_0^{(0)}}\bra{\Psi_I^{(0)}}$, $\hat{h}_{\nu}^{(0)\dagger} = \ket{\Psi_J^{(0)}}\bra{\Psi_0^{(0)}}$, with $I$ and $J \ne 0$) for the reference state $\ket{\Psi_0^{(0)}}$ with the Hermitian QD-NEVPT2 effective Hamiltonian in \cref{eq:eff_H_sym_0,eq:eff_H_sym_1,eq:eff_H_sym_2}.
The contributions to $M_{\mu\nu}$ can be simplified in the following form:
\begin{align}
	\label{eq:M_EE_0}
	{M}^{(0)}_{IJ} &= (E^{(0)}_I - E^{(0)}_0) \delta_{IJ}   \\
	\label{eq:M_EE_2}
	{M}^{(2)}_{IJ} &= 
	\frac{1}{2} \langle{\Psi^{(0)}_{I}}|\hat{{V}}|{\tilde{\Psi}^{(1)}_J}\rangle 
	+ \frac{1}{2} \langle{\tilde{\Psi}^{(1)}_{I}}|\hat{{V}}|{\Psi^{(0)}_J}\rangle  \notag \\
	&+ \frac{1}{2} \langle{\Psi^{(0)}_{I}}|\tilde{H}^{(1)}|{\tilde{\Psi}^{(1)}_J}\rangle 
	+ \frac{1}{2} \langle{\tilde{\Psi}^{(1)}_{I}}|\tilde{H}^{(1)}|{\Psi^{(0)}_J}\rangle 
	- E^{(2)}_0 \delta_{IJ}
\end{align}
where $E^{(0)}_0$ and $E^{(2)}_0$ are the reference CASSCF and pc-NEVPT2 correlation energies (\cref{eq:corr_energy_2}), respectively, $\tilde{H}^{(1)}$ is defined in \cref{eq:H_bch_1}, and $\ket{\tilde{\Psi}^{(1)}_I} = \hat{T}^{(1)} \ket{{\Psi}^{(0)}_I}$ with the amplitudes of $\hat{T}^{(1)}$ determined for the reference state $\ket{\Psi_0^{(0)}}$ (as opposed to the amplitudes of $\ket{{\Psi}^{(1)}_I}$ determined for $\ket{{\Psi}^{(0)}_I}$ as defined in \cref{eq:1st_order_wfn_contracted}).

Comparing \cref{eq:M_EE_0,eq:M_EE_2} to \cref{eq:eff_H_sym_0,eq:eff_H_sym_2}, we observe that at the zeroth order the EE-MR-ADC(2) and QD-NEVPT2 effective Hamiltonians differ only in the constant shift by the CASSCF reference energy ($-E^{(0)}_0\delta_{IJ}$) due to the fact that EE-MR-ADC(2) directly approximates the excitation energies rather than the energies of individual electronic states. 
The differences in the second-order contributions stem from two sources.

First, EE-MR-ADC(2) incorporates the second-order correction to the reference energy $E^{(2)}_0$ neglecting the coupling between the perturbed reference wavefunction $\ket{\Psi_0^{(1)}}$ and the excited states $\ket{\Psi_I^{(0)}}$ and $\ket{\Psi_J^{(0)}}$ while this coupling is explicitly accounted for in the diagonalization of effective Hamiltonian in QD-NEVPT2.
For this reason, EE-MR-ADC(2) is unable to correctly describe the avoided crossing between the reference and excited states as opposed to QD-NEVPT2, which incorporates the interaction between these states.

Second, the EE-MR-ADC(2) effective Hamiltonian is defined in terms of the reference state amplitudes for $\ket{\tilde{\Psi}^{(1)}_I}$ and $\tilde{H}^{(1)}$ as opposed to the state-specific amplitudes for $\ket{{\Psi}^{(1)}_I}$ in QD-NEVPT2.
This modifies the form of the first two terms in \cref{eq:M_EE_2} and gives rise to the appearance of two new contributions from $\tilde{H}^{(1)}$ ({\it cf.}\@ \cref{eq:eff_H_sym_2}).
It can be shown that $\langle{{\Psi}^{(1)}_{I}}|\tilde{H}^{(1)}|{\Psi^{(0)}_I}\rangle = \langle{{\Psi}^{(0)}_{I}}|\tilde{H}^{(1)}|{\Psi^{(1)}_I}\rangle  = 0$ if $\tilde{H}^{(1)}$ is defined in terms of the amplitudes of $\ket{\Psi^{(1)}_I}$ and, thus, these contributions do not need to be accounted for in the QD-NEVPT2 effective Hamiltonian. 
Importantly, the QD-NEVPT2 method does not incorporate the contributions from the off-diagonal matrix elements   $\langle{{\Psi}^{(1)}_{I}}|\tilde{H}^{(1)}|{\Psi^{(0)}_J}\rangle$ and $\langle{{\Psi}^{(0)}_{I}}|\tilde{H}^{(1)}|{\Psi^{(1)}_J}\rangle$ ($I\ne J$) in its effective Hamiltonian, which are in general non-zero.
The lack of these terms gives rise to the model space invariance of QD-NEVPT2 as discussed in Ref.\@ \citenum{Lang:2020p014109}.

{\it Computational cost and scaling.} 
MR-ADC(2) and MR-ADC(2)-X require calculating up to 4-RDM with the $\mathcal{O}(N_{act}^8 N_{det})$ computational scaling where $N_{act}$ and $N_{det}$ are the numbers of active-space orbitals and Slater determinants, respectively. 
Similar to NEVPT2 and QD-NEVPT2, both methods have been implemented using the intermediate factorization techniques (\cref{sec:nevpt:high_order_rdms}) that avoid 4-RDM and lower the scaling to $\mathcal{O}(N_{act}^6 N_{det})$.
For fixed $N_{act}$, EE/EA/IP-MR-ADC(2), EA/IP-MR-ADC(2)-X, NEVPT2, and QD-NEVPT2 have the same $\mathcal{O}(N^5)$ computational scaling with the basis set size $N$. 
The computational cost of EE-MR-ADC(2)-X method scales as $\mathcal{O}(N^6)$.

{\it Excited-state properties.} 
The MR-ADC methods offer direct and straightforward access to a wide range of transition properties, such as spectroscopic factors, oscillator strengths, Dyson orbitals, and densities of states. 
In addition, state-specific properties can be computed by evaluating the reduced density matrices of correlated excited states.
As discussed in \cref{sec:nevpt:analytic_gradients_properties}, NEVPT and QD-NEVPT have been used to compute state-specific properties, such as equilibrium geometries, zero-field splitting, and g-tensors. 

\section{Other approaches based on the Dyall Hamiltonian}
\label{sec:other}

The Dyall Hamiltonian has been widely used outside the domains of NEVPT and MR-ADC. 
\cref{tab:dyall_ref}  lists the theoretical approaches that employ the Dyall Hamiltonian as the zeroth-order Hamiltonian for a variety of applications, such as calculating correlation energies, relative energies between electronic states, and triple excitation corrections.
Additionally, the Dyall Hamiltonian has been employed to treat linear dependencies and improve convergence in canonical transformation theory \cite{Neuscamman:2010p024106,Neuscamman:2010p231}.
The use of the Dyall Hamiltonian has been also considered in multireference driven similarity renormalization group \cite{Li:2017p124132} and in methods based on adiabatic connection \cite{Matousek:2023p054105}. 

\begin{table*}[t]
	\caption{
		Theoretical approaches that employ the Dyall Hamiltonian outside the domains of NEVPT and MR-ADC.
	}
	\label{tab:dyall_ref}
	\setstretch{1}
	\scriptsize
	\centering
	\begin{threeparttable}
		\begin{tabular}{lc}
			\hline\hline
			Description			&  Reference	\\ 
			\hline
			Perturbative triples correction in internally contracted multireference coupled cluster theory& \citenum{Hanauer:2012p204107,Kohn:2020pe1743889} \\	
			Jeziorski--Monkhorst uncontracted multireference perturbation theory 			& \citenum{Giner:2017p224108} \\	
			Static-dynamic-static multistate multireference second-order perturbation theory 			& \citenum{Lei:2017p2696} \\	
			Second-order dynamic correlation dressed complete active space method 			& \citenum{Pathak:2017p234109,Lang:2020p1025} \\	
			Geminal perturbation theory based on the unrestricted Hartree--Fock wavefunction 			& \citenum{Foldvari:2019p034103} \\	
			Super-CI approach for the orbital optimization in CASSCF  			& \citenum{Kollmar:2019p399} \\	
			Second-order approximate internally contracted multireference coupled-cluster theory			& \citenum{Kohn:2019p041106,Aoto:2019p2291} \\	
			CASPT2 with modified zeroth-order Hamiltonian 			& \citenum{Kollmar:2020p214110} \\	
			\hline\hline
		\end{tabular}
	\end{threeparttable}
\end{table*}

\section{Summary and outlook}
\label{sec:summary}

In this review, we presented an overview of multireference perturbation theories based on the Dyall Hamiltonian with a particular focus on $N$-electron valence perturbation theory (NEVPT), its quasidegenerate formulation (QD-NEVPT), and multireference algebraic diagrammatic construction theory (MR-ADC).
The concept of Dyall Hamiltonian \cite{Dyall:1995p4909} contributed significantly  to the development of intruder-free multireference theories, establishing new theoretical approaches that capture static and dynamic electron correlation for molecular systems with large active spaces and basis sets. 

There are many opportunities in advancing the development of theories based on the Dyall Hamiltonian further. 
Despite active research, the implementations of NEVPT and QD-NEVPT are currently limited to the calculations with $\lesssim$ 30 active orbitals.
Overcoming this limitation requires implementing  NEVPT and QD-NEVPT without three- and four-particle reduced density matrices and developing more efficient approaches for capturing static correlation in the active space, as exemplified by the recent work on NEVPT with reference wavefunctions from quantum computers \cite{Tammaro:2023p817}.
Additional developments are necessary to expand the range of molecular properties that can be computed using NEVPT and QD-NEVPT.

The development of MR-ADC is still in its infancy. 
Efficient implementations of MR-ADC methods will enable the reliable simulations of electronic spectra for realistic multireference systems, such as transition metal complexes and large organic chromophores. 
Another exciting area of research is the development of multireference methods for periodic two- and three-dimensional materials. 
These chemical systems present new challenges, such as the treatment of translational symmetry, screening effects, and phase transitions. 

Multireference calculations of large extended and molecular systems using the Dyall Hamiltonian will require developing novel approaches to reduce computational costs.
In this context, the use of machine learning and embedding techniques appears to be quite promising. 
Further progress in this direction should involve combining multireference electronic structure methods with approaches for capturing vibrational and phonon effects, ab initio molecular dynamics, and sampling of different configurations on free energy surfaces. 

Research along these directions will improve the understanding of chemical systems with complicated electronic structure from first principles of quantum mechanics.

\section{Acknowledgments}
A.Y.S.\@ acknowledges the support from National Science Foundation, under Grant No.\@ CHE-2044648.


\bibliographystyle{jcp} 

\end{document}